\documentclass[%
 reprint,
 amsmath,amssymb,
 aps,
]{revtex4-2}

\usepackage{graphicx}
\usepackage{dcolumn}
\usepackage{bm}
\usepackage[
    colorlinks=true,
    linkcolor=blue,
    urlcolor=blue,
    citecolor=blue]{hyperref}

\usepackage{xcolor}

\def\prn#1{{\left(#1\right)}}

\def\bs{\boldsymbol}

\usepackage[normalem]{ulem}

\usepackage{tikz}
\usetikzlibrary{shapes}

\usepackage{array, multirow, boldline}
\usepackage{booktabs}

\usepackage[]{ezedits}
\defineEdit{DK}{\color[rgb]{.1,.1,.8}}{\color[rgb]{0.1,.1,.8}}

\defineEdit{TL}{\color[rgb]{.8,.1,.1}}{\color[rgb]{0.8,.1,.1}}

\begin{document}


\title{Optimization of nuclear polarization in an alkali-noble gas comagnetometer}

\author{Emmanuel Klinger}
\thanks{present address: Universit\'e de Franche-Comt\'e, SupMicroTech-ENSMM, UMR 6174 CNRS, institut FEMTO-ST, 25000 Besan\c{c}on, France}
\email{emmanuel.klinger@femto-st.fr}
\affiliation{Johannes Gutenberg-Universit\"at Mainz, 55128 Mainz, Germany}
 \affiliation{Helmholtz-Institut Mainz, GSI Helmholtzzentrum f{\"u}r Schwerionenforschung, 55128 Mainz, Germany}
 
\author{Tianhao Liu}
\affiliation{Johannes Gutenberg-Universit\"at Mainz, 55128 Mainz, Germany}
\affiliation{Helmholtz-Institut Mainz, GSI Helmholtzzentrum f{\"u}r Schwerionenforschung, 55128 Mainz, Germany}

\author{Mikhail Padniuk}
\affiliation{Marian Smoluchowski Institute of Physics, Jagiellonian University in Krakow, Łojasiewicza 11, 30-348, Krakow, Poland}

\author{Martin Engler}
\affiliation{Johannes Gutenberg-Universit\"at Mainz, 55128 Mainz, Germany}
\affiliation{Helmholtz-Institut Mainz, GSI Helmholtzzentrum f{\"u}r Schwerionenforschung, 55128 Mainz, Germany}

\author{Thomas Kornack}
\affiliation{Twinleaf LLC, 300 Deer Creek Drive, Plainsboro, NJ 08536, USA}

\author{Szymon Pustelny}
\affiliation{Marian Smoluchowski Institute of Physics, Jagiellonian University in Krakow, Łojasiewicza 11, 30-348, Krakow, Poland}

\author{Derek F. Jackson Kimball}
\affiliation{Department of Physics, California State University – East Bay, Hayward, CA 94542, USA}

\author{Dmitry Budker}
\affiliation{Johannes Gutenberg-Universit\"at Mainz, 55128 Mainz, Germany}
\affiliation{Helmholtz-Institut Mainz, GSI Helmholtzzentrum f{\"u}r Schwerionenforschung, 55128 Mainz, Germany}
\affiliation{Department of Physics, University of California, Berkeley, CA 94720, USA}

\author{Arne Wickenbrock}
\affiliation{Johannes Gutenberg-Universit\"at Mainz, 55128 Mainz, Germany}
\affiliation{Helmholtz-Institut Mainz, GSI Helmholtzzentrum f{\"u}r Schwerionenforschung, 55128 Mainz, Germany}

\begin{abstract}

Self-compensated comagnetometers, employing overlapping samples of spin-polarized alkali and noble gases (for example K-$^3$He) are promising sensors for exotic beyond-the-standard-model fields and high-precision metrology such as rotation sensing.  When the comagnetometer operates in the so-called self-compensated regime, the effective field, originating from contact interactions between the alkali valence electrons and the noble-gas nuclei, is compensated with an applied magnetic field. When the comagnetometer begins operation in a given magnetic field, spin-exchange optical pumping establishes equilibrium between the alkali electron-spin polarization and the nuclear-spin polarization. Subsequently, when the magnetic field is tuned to the compensation point, the spin polarization is brought out of the equilibrium conditions.  This causes a practical issue for long measurement times. 
We report on a novel method for closed-loop control of the compensation field. This method allows optimization of the operating parameters, especially magnetic field gradients, in spite of the inherently slow (hours to days) dynamics of the system. With the optimization, higher stable nuclear polarization, longer relaxation times and stronger electron-nuclear coupling are achieved which is useful for nuclear-spin-based quantum memory, spin amplifiers and gyroscopes.  The optimized sensor demonstrates a sensitivity comparable to the best previous comagnetometer but with four times lower noble gas density. 
This paves the way for applications in both fundamental and applied science.
\end{abstract}

\maketitle

\section{Introduction}

Over the past 20 years, vapor-cell-based atomic sensors \cite{kitchingsensors} have received growing attention
due to the possibilities of 
miniaturization and low power consumption \cite{kitching2018chip}, features, which compare favorably to cryogenically cooled sensors such as superconducting interferometers (SQUIDs).

In a gas containing alkali atoms and polarized noble gas atoms, the electrons of the alkalis feel an effective magnetic field resulting from contact interactions between the electrons and the noble-gas nuclei. An applied external magnetic field can be tuned to cancel this effective field, then such a comagnetometer operates at the so-called compensation point. Under these conditions, the polarized noble-gas nuclei can adiabatically follow a slowly changing magnetic field. This renders the device insensitive to transverse magnetic fields that vary slower than the response time of the system, typically, several hundreds of milliseconds \cite{kornack2002dynamics}. 
Another advantage of operating near the compensation field is that the alkali vapor may be brought into the spin-exchange-relaxation-free (SERF) regime \cite{Happer_Tam_PhysRevA_1977},
if the frequency of the alkali-spin Larmor precession is much lower than the rate of spin-exchange collisions. Operation in the SERF regime significantly improves the sensitivity of the comagnetometer \cite{allred2002high-sensitivity} to nonmagnetic perturbations. 

While designed to be insensitive to magnetic fields, self-compensated comagnetometers are exquisitely sensitive to nonmagnetic interactions. In the self-compensation regime the noble-gas magnetisation follows changes in the low-frequency external magnetic field, which protects the alkali-metal polarization from the magnetic-field perturbation \cite{kornack2002dynamics,padniuk2022response}. However, this protection does not extend to nonmagnetic perturbations (e.g. rotation or exotic spin couplings), allowing the measurement of such perturbations. First demonstrated in the early 2000s \cite{kornack2002dynamics}, these devices were extensively improved in follow-up studies. It was shown that the effects of light shifts and radiation trapping can be minimized if the probed alkali species is polarized by spin-exchange optical pumping (hybrid pumping) \cite{romalis2010hybrid}. Recently, the response of the system to a low-frequency field modulation was explored \cite{lu2020nuclear}, which can be used for \textit{in-situ} characterization of the comagnetometer at the compensation point. 
Noble-gas comagnetometers have proven to be sensitive gyroscopes \cite{kornack2005nuclear,jiang2018parametrically,Liu2022Comag,Liang2022Biaxial,wei2022ultrasensitive} and powerful tools for exotic physics searches \cite{vasilakis2009limits,terrano2021comag,blocholdcomag,padniuk2022response,wei2022ultrasensitive}. Due to the long coherence time of noble-gas spins and the ability to optically manipulate alkali-atom spins, comagnetometers have found applications in quantum memory assisted by spin exchange \cite{katz2022quantum,shaham2022strong,katz2022optical}. 

Although extremely sensitive, self-compensated comagnetometers are often difficult to operate for a long period of time (days) because of the need of frequent field zeroing, alignment of laser beams, etc. 
One way to improve the stability of the system 
is to use a high noble-gas pressure (for example 12\,amg was used in Ref.\,\cite{vasilakis2009limits}, where  $1\,\text{amg}=2.69\times10^{19}\,$cm$^{-3}$). Although this approach is applicable for individual experiments, it poses challenges of producing and handling multiple gas-containing vapor cells on larger scales 
and thus acts as a bottleneck for wide usability in applications. 

Here, we report on a novel method for following, in real time, the build-up of nuclear magnetization through spin-exchange optical pumping. This enables reaching stationary working conditions, where the effective magnetic field experienced by alkali atoms is approximately zero.
A ${^3}$He-$^{39}$K-$^{87}$Rb comagnetometer is successfully operated at equilibrium nuclear polarization above 100\,nT with a cell pressure below 3\,amg \footnote{In our work, the nuclear polarization in the self-compensated regime was above 3\%. For comparison, the nuclear polarization in Ref.\,\cite{vasilakis2009limits} was comparable however with four times the noble gas density.} without frequent field zeroing. This is enabled by precise compensation of the field gradients inside the magnetic shield. In Sec.\,\ref{sec:equilirbium}, we review the effect of magnetic field gradients on the comagnetometer, showing that for a given cell, equilibrium nuclear magnetization can only be achieved at high pressure or low field gradients. In Sec.\,\ref{sec:experiment}, we present the experimental setup and show the measured effect of field gradients on the transverse 
and longitudinal 
relaxation rates and the evolution of nuclear magnetization over time. Polarization dynamics is further discussed in Sec.\,\ref{sec:polDyna} and the conclusions are drawn in Sec.\,\ref{sec:Conclusion}.

\section{Comagnetometer theory}\label{sec:equilirbium}

We are interested in understanding how field gradients affect the equilibrium nuclear magnetization. Consider an ensemble of polarized electronic ($^{39}$K) and nuclear ($^3$He) spins enclosed in a perfectly spherical cell. 
The dynamics of the longitudinal component (along the light propagation direction, $z$) of the nuclear polarization follows\,\cite{kornack2002dynamics}
\begin{equation} \label{eq:dynamic_pumping}
\frac{d}{dt}P^n(t)={R^{ne}_\text{se}}(P^e-P^n)-\frac{1}{T_1^n}P^n\,,
\end{equation}
where $P^n(t)$ is the time-dependent fractional $^3$He polarization, for which the steady-state value is determined as
\begin{equation} \label{eq:nuc_pol}
P^n(t\rightarrow\infty)=\frac{R^{ne}_\text{se}}{1/T_1^n+R^{ne}_\text{se}}P^e\,.
\end{equation}
Here $P^e$ is the electronic spin polarization (in typical high sensitivity experiments, $P^e \approx 50\%$ \cite{shah2009spin}), $R^{ne}_\text{se}=\mathcal{N}_\text{e}\sigma^{ne}_{\text{se}}v$ is the spin-exchange rate for a noble gas atom, $\sigma^{ne}_{\text{se}}$ is the spin-exchange cross-section, $v$ is the characteristic relative velocity in alkali-noble-gas collisions, $\mathcal{N}_\text{e}$ is the concentration of alkali atoms,  and $T_1^n$ is the longitudinal relaxation rate of nuclear spins.



The main contribution to the relaxation time $T_1^n$ in many experiments comes from magnetic field inhomogeneities throughout the cell. In this case and for sufficiently high pressures \cite{schearer1965,Cates1988inhomogeneity},
\begin{equation} \label{eq:T1_inh}
\frac{1}{T^n_1}=D_{n}\frac{|\nabla B_\perp|^2}{B_z^2},
\end{equation}
where $B_z$ is the leading field seen by the nuclear spins and the transverse field gradients are characterized by $|\nabla B_\perp|^2 = |\nabla B_x|^2 + |\nabla B_y|^2$. There are various contributions to the field gradient, including the residual gradients from the magnetic shields, the field coils, as well as gradients due to the polarized spins (in our case dominated by nuclear spins). Note that the gradient due to polarized spins only occurs when the cell is not spherical. 
This gradient is proportional to $P^n \mathcal{N}_n$ and cannot be fully compensated by first-order gradient coils. Here we describe this effect using a phenomenological factor $\alpha$, accounting for the asphericity of the cell, making the overall gradient 
\begin{equation} \label{eq:Grad}
 |\nabla B_\perp|^2=|\nabla B_{\perp,0}|^2 + |\alpha P^n \mathcal{N}_n|^2\,,
\end{equation}
with  $|\nabla B_{\perp,0}|^2 $ representing the first-order gradient that can be zeroed by gradient coils. 

The noble gas self-diffusion coefficient $D_{n}$ is a function of the temperature $\Theta$ and the noble gas density $\mathcal{N}_{n}$. It can be approximated by
\begin{equation}\label{eq:Diffusion}
D_{n}=D_0~\frac{\sqrt{\Theta/273\,\textrm{K}}}{\mathcal{N}_{n}/1\,\textrm{amg}}\,.
\end{equation} 
For $D_0=1.65\,\text{cm}^2/\text{s}$ in the case of $^3$He \cite{HeDiffusion1974}, $\Theta=458\,\text{K}$ ($185^\circ$C), and $\mathcal{N}_{n} \approx 3\,\text{amg}$, the self-diffusion coefficient is roughly $0.71\,\text{cm}^2/\text{s}$. 


\subsection{Compensation point}


In the comagnetometer, the principal mechanism via which nuclear and electronic spins couple to each other is spin-exchange collisions. Under particular conditions it turns out that this coupling can result in a suppression of sensitivity to slowly changing transverse magnetic fields.
This suppression is maximal when the applied longitudinal magnetic field is tuned to the so-called compensation point $B_c$, where the coupling of nuclear and electronic spins is the strongest \cite{kornack2002dynamics},

\begin{equation}\label{eq:comp-point-all-contrib}
   B_c=- B_n - B_e = -\lambda M_n P^n - \lambda M_e P^e.
\end{equation}
Here $\lambda=2\kappa_0\mu_0/3$ \footnote{In CGS units: $\lambda=8\pi\kappa_0/3$.} with $\mu_0$ being the vacuum permeability and $\kappa_0$ being the spin-exchange enhancement factor due to the overlap of the alkali electron wave function and the nucleus of the noble gas \cite{baranga1998}; $M=\mu\,\mathcal{N}$ corresponds to a fully polarized sample with atoms of magnetic moment $\mu$ and density $\mathcal{N}$.

As an aside, note that the transverse noble-gas nuclear spin damping rate, hence the comagnetometer bandwidth, is maximal if the applied longitudinal field matches \cite{lee2019new}:
\begin{equation} \label{eq:dampingCompPoint}
    B_d= - B_n + B_e Q\frac{\gamma_n}{\gamma_e},
\end{equation}
where $Q(P^e)=4\left[2-4/(3+P^e\,^2)\right]^{-1}$ is the electron slowing down factor for $^{39}$K \cite{savukov2005effects}; $\gamma_e$ and $\gamma_n$ are the electronic and nuclear spin gyromagnetic ratios, respectively. This point is often referred to as the fast-damping field, which is better suited for high-bandwidth rotation measurements.

Under our experimental conditions, electronic spin magnetization $ B_e $ is small ($\sim 2$\,nT)  compared to nuclear spin magnetization $ B_n $ ($\sim100$\,nT) \cite{kornack2005nuclear}, leading to
\begin{equation} \label{eq:simpleCompPoint}
B_c \approx B_d \approx - B_n= -\lambda M_n P^n.
\end{equation}
At this point, alkali spins are to first-order insensitive  to transverse magnetic fields but are still sensitive to non-magnetic interactions \cite{kornack2005nuclear}.


\begin{figure}
    \centering
    \includegraphics[scale=0.91]{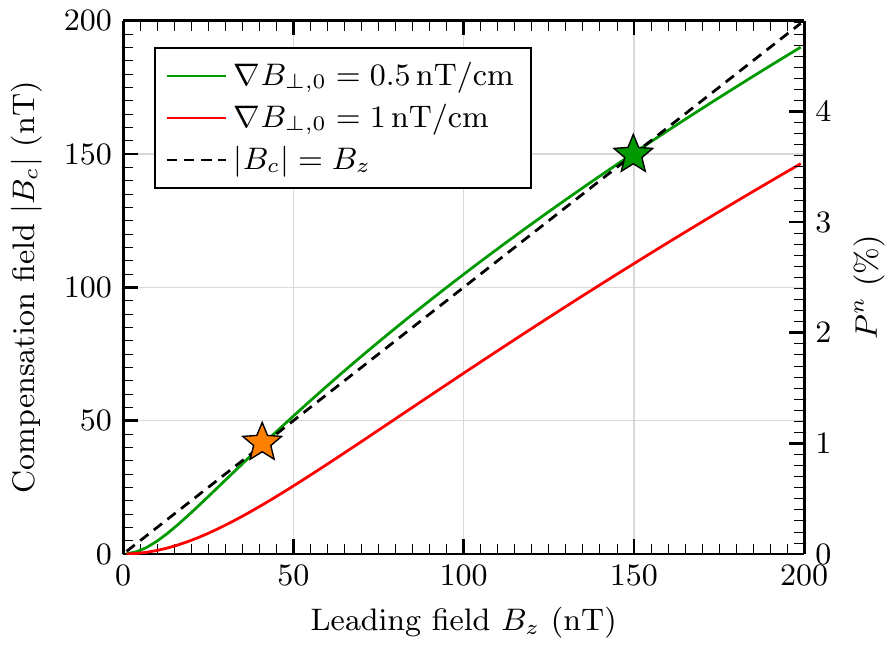}
    \caption{Absolute value of the steady state compensation field as a function of leading field $B_z$ at high gradient (1\,nT/cm, red solid line) and low gradient (0.5\,nT/cm, green solid line). When the system is brought to such a field, it will generally self-adjust over time. However, the intersections of the green line with the dashed black line ($|B_c|=B_z$), marked with stars, correspond to equilibrium compensation points. Here, we make a distinction between the lower stationary compensation point (orange star) and the upper one (green star). These two points correspond to the roots of Eq.\,\eqref{eq:quadBzEq}; however, the system is stable near the upper point and unstable near the lower one. In the simulation, the asphericity factor is $\alpha \mathcal{N}_n = 25$\,nT/cm and the spin-exchange polarization rate is $R^{ne}_\text{se}=4.5\times 10^{-6}$\,s$^{-1}$. 
 }
    \label{fig:theory}
\end{figure}

\subsection{Equilibrium condition}\label{sec:equilibrium-condition}
Suppose the comagnetometer has reached steady-state conditions for a given $B_z$, which is not the compensation point. Generally, if the device is brought to the compensation point by setting $B_z=B_c$, we will find that the system is no longer in equilibrium: the nuclear spin polarization will change over time. This can be seen, for instance, from the combination of Eqs.~\eqref{eq:nuc_pol} and \eqref{eq:T1_inh}, which shows that the steady-state polarization depends on $B_z$ (if the transverse gradient  $\nabla B_\perp$ is not proportional to $B_z$). Consequently, $B_c$ changes because $B_c \approx -B_n = -\lambda M_n P^n$, and the system is no longer at the compensation point.

If we wish to keep the system at the compensation point over an extended measurement time, it is possible to experimentally track it and lock the leading field such that $B_z=B_c$ \cite{jiang2019closed,tang2022design}. In many cases, in closed-loop operation, however, the nuclear polarization will gradually decrease to zero, degrading the sensitivity of the comagnetometer. 

It turns out, there exist special values of $B_z$ where the system is in stable equilibrium at the compensation point. Here, the polarization does not change with time as the equilibrium polarization corresponds to the  compensation point.
From Eqs.~\eqref{eq:nuc_pol}, \eqref{eq:T1_inh} and \eqref{eq:simpleCompPoint}, one finds an equation for the leading field corresponding to the stationary system
\begin{equation}\label{eq:quadBzEq}
   B_z^2+ \lambda M_n P^e B_z+ \frac{D_n}{ R_{\text{se}}^{ne}}|\nabla B_\perp|^2=0\,.
\end{equation}

A solution for $B_z$ exists only if
\begin{equation}
    \left(\lambda M_n P^e \right)^2-4\frac{D_n}{R_{\text{se}}^{ne}}|\nabla B_\perp|^2\geqslant0\,, 
\end{equation}
that is
\begin{equation}
 \sqrt{\frac{R_{\text{se}}^{ne}}{4D_n}} \lambda M_n P^e \geqslant |\nabla B_\perp|\,.
\label{eq:gradientCondition}
\end{equation}

This illustrates why high noble-gas pressures are desirable. Indeed, noting that the diffusion coefficient $D_n$ is inversely proportional to $\mathcal{N}_n$, the magnitude of tolerable gradients scales as $\mathcal{N}_n^{3/2}$. Because $P^e$ is limited to unity, and $\lambda$ is a constant, the only ways to ensure the comagnetometer operates at the equilibrium compensation point is to use high nuclear spin density and/or more accurately zero the gradients. 

In Fig.\,\ref{fig:theory}, one sees that, for a constant gradient $|\nabla B_\perp|$, the nuclear polarization, and hence the compensation field $B_c$, 
grows with external magnetic field $B_z$. In this figure, stationary
compensation points can be found by looking for the intersection of the $B_c$ curve with the diagonal (black dashed) line. In the case of $|\nabla B_\perp|=0.5$\,nT/cm (green solid line), these stationary points are indicated with orange and green stars. However, when $|\nabla B_\perp|=1$\,nT/cm (red solid line), no stationary compensation point exists. 

In practice, part of the gradient may be related to the coils generating the leading field, therefore the field gradient may change with the $B_z$ field. This effect is not included in the results presented in Fig.\,\ref{fig:theory}. 


\section{Gradient compensation }\label{sec:experiment}
\subsection{Experimental setup}
\begin{figure}[htb]
    \centering
    \includegraphics[width=0.48\textwidth]{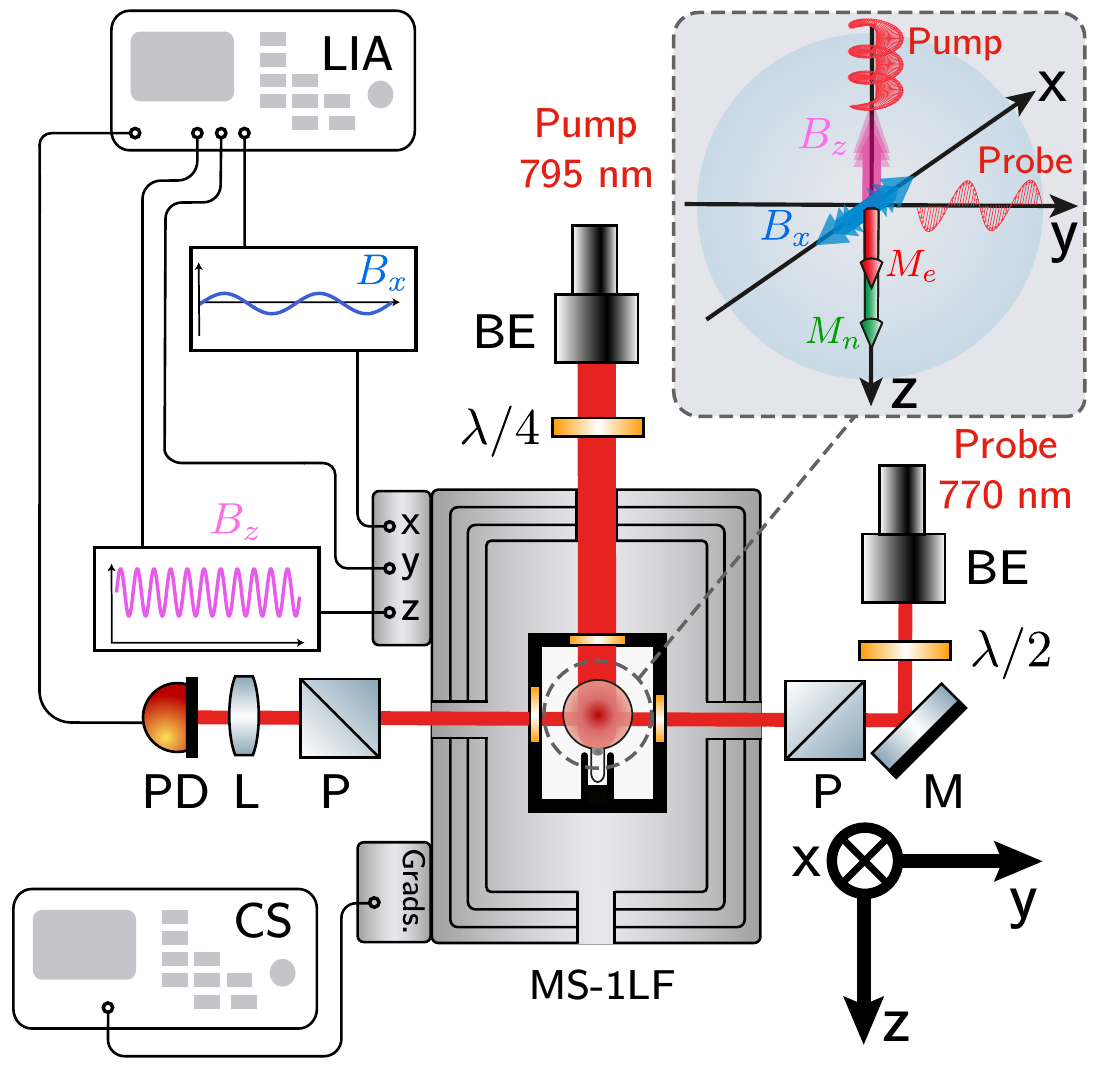}
    \caption{Sketch of the experimental setup. LIA -- lock-in amplifier, PD -- photodetector, L -- lens, CS -- low-noise current source, P -- polarizer, M -- mirror, BE -- beam expander, $\lambda/4$ -- quarter-wave plate, $\lambda/2$ -- half-wave plate, MS1-LF -- magnetic shield. The inset shows the directions and polarizations of the laser beams, magnetic field modulations and the directions of the generated electronic and nuclear magnetization. The $B_x$ modulation is only used for closed-loop control of the compensation point and $^3$He magnetization measurement, see Sec.\,\ref{sec:opti-comp-point}. }
    \label{fig:setup}
\end{figure}

The experimental setup is sketched in Fig.\,\ref{fig:setup}. We use a 20-mm-diameter spherical cell filled with 3\,amg of $^3$He and 50\,torr of N$_2$. 
The cell has a sidearm loaded with a drop of alkali-metal mixture, 1\%\,$^{87}$Rb and 99\%\,K molar fractions. The vapor cell is placed in a Tw\^inleaf MS-1LF magnetic shield. To minimize the effect of cell asphericity \cite{kornack2002dynamics,romalis2014comment}, the comagnetometer was mounted up-right ($z$-axis along gravity) so that the cell sidearm is plugged by the liquid droplet of alkali metals. Note that other orientations of the stem are sometimes found to minimize the effects of asphericity \cite{Terrano2019PRA}. Copper wires (not shown in Fig.\,\ref{fig:setup}) are looped around the cylindrical layers for degaussing of the shield and 
its content. We note that the cell was demagnetized with a commercial demagnetizer prior to being mounted in the oven. Self-magnetization of the cell, e.g. due to ferromagnetic impurities in the glass, is known to affect the usual quadratic dependence of $T^n_1$ to $B_z$, see Eq.\,\eqref{eq:T1_inh}. In the presence of self magnetization of the cell, $T^n_1$ will be proportional $B_z$ instead \cite{brown2011new}. The working temperature of $185^\circ$C is achieved with AC resistive heating. At this temperature, the ratio of Rb to K concentrations in the vapor is 3:97, which was measured by fitting the transmission spectrum of Rb D$_2$ and K  D$_1$ lines.

Rubidium atoms are optically pumped by 70\,mW of circularly-polarized light from a Toptica TA Pro laser in resonance with the Rb D$_1$ line. For uniform pumping over the cell, the beam is expanded to 20-mm diameter. Potassium (and helium) spins are pumped by spin-exchange collisions with Rb. The comagnetometer readout is realized by monitoring the polarization rotation of a  linearly polarized $\approx16\,\text{mW}/\text{cm}^2$ (average intensity of 7\,mW and beam diameter of 7.5\,mm) probe beam (Toptica DL Pro) detuned about 0.5\,nm toward shorter wavelength from the K D$_1$ line. Because K atoms are pumped by spin-exchange optical pumping, the SERF magnetometer is much less sensitive to light shifts of the pump beam \cite{romalis2010hybrid}. Both pump and probe beams are guided to the setup with optical fibers. No active stabilization of the lasers is performed apart from that of temperature and current of the diode lasers.

To perform low-noise detection of the response to perturbations along the $y$-axis, the $B_z$ field is modulated with a sine wave (800\,Hz, 35\,nT peak-to-peak) \cite{zhimin2006parametric,jiang2018parametrically} and the comagnetometer signal is analyzed with a lock-in amplifier. Details on the parametric modulation and chosen parameters are given in Appendix\,\ref{sec:appendI-serf}.
In typical SERF magnetometers, the lock-in detection is achieved by modulating the incident or output probe-beam polarization with photoelastic or Faraday modulators. Using magnetic field modulation instead is helpful to improve the compactness as well as lowering the cost but the SERF resonance is then slightly broadened because of the modulation field, see Appendix\,\ref{sec:appendI-serf}.

After zeroing the magnetic field in the cell and optimizing the pump and probe beam frequencies and powers, the width of the SERF resonance was reduced to 8\,nT, leading to a sensitivity to $B_y$ field better than $10\,\text{fT}/\sqrt{\text{Hz}}$ at 20\,Hz, limited by photon shot noise. 
This characterization was done prior to the gradient optimization and operating the system in the self-compensated regime. 

\subsection{Gradient optimization}
From Eq.\,\eqref{eq:T1_inh}, one sees that the key to maximizing the nuclear polarization is reduction of transverse gradients
\begin{equation}\label{eq:orthoGradComponents}
\begin{split}
|\nabla B_\perp|^2=\prn{\partial_xB_x}^2&+\prn{\partial_yB_x}^2+\prn{\partial_zB_x}^2 \\&+\prn{\partial_xB_y}^2+\prn{\partial_yB_y}^2\\
&+\prn{\partial_zB_y}^2\,.
\end{split}
\end{equation}


The Tw\^inleaf MS-1LF magnetic shield provides 5 independent first-order gradient coils in a built-in flexible printed circuit board, allowing the shimming of the field gradient tensor. To zero transverse gradients, we use a zeroing procedure with two steps. The first step, as we will explain below (Sec.\,\ref{sec:T2method}), is to use $T^n_2$ as an indicator to zero the gradients of the longitudinal field $B_z$. Then, because $\bs{\nabla} \times \bs{B} = 0$, the two $T^n_1$-relevant components appearing in Eq.~\eqref{eq:orthoGradComponents}, $\partial_xB_z=\partial_zB_x$ and $\partial_yB_z=\partial_zB_y$, are also nulled. The second step is to zero the other remaining independent components using the change of nuclear polarization as an indicator, see Sec.\,\ref{sec:opti-comp-point}.

\subsubsection{Optimization at low nuclear polarization}\label{sec:T2method}
At first, the gradients of the longitudinal field $B_z$ are optimized away from the compensation point, at low nuclear polarization. The transverse decay rate of $^3$He spins is known to be a function of the diffusion across $z$-field gradients \cite{Cates1988inhomogeneity}
\begin{equation}
    \frac{1}{T^n_2}=\frac{8a^4\gamma_n^2|\nabla B_z|^2}{175 D_{n}}\,,
    \label{eq:T2LongitudinalGrad}
\end{equation}
where $a$ is the cell radius and $\gamma_n$ is the nuclear gyromagnetic ratio.

To accurately zero $|\nabla B_z|$, the following procedure is used: \textit{(i)} Helium is polarized for 5\,min with $B_z=100$\,nT. \textit{(ii)}  The $B_y$ field is incremented by 5.5\,nT to adiabatically tip the helium spins away from the $z$-axis. \textit{(iii)} After turning the $B_y$ field off, the spins are left to precess. The precession is measured for 5\,s by monitoring probe beam before the pump beam is turned off (the weak probe beam is left on). Then the spins precess in the dark for 5\,min before the pump beam is turned on to measure the precession signal again. The two precession amplitudes are compared to estimate $T^n_2$. \textit{(iv)} Nuclear polarization is destroyed by applying a $\partial_zB_z$ gradient of the order of 30\,nT/cm along the $z$-axis for 5\,s. This is long enough to depolarize He spins, for details, see Appendix\,\ref{sec:appendI-depol}.
 The last step is important to make sure all measurements are realized at the same nuclear polarization.
 
\begin{figure}[htb]
    \centering
    \includegraphics[scale=0.92]{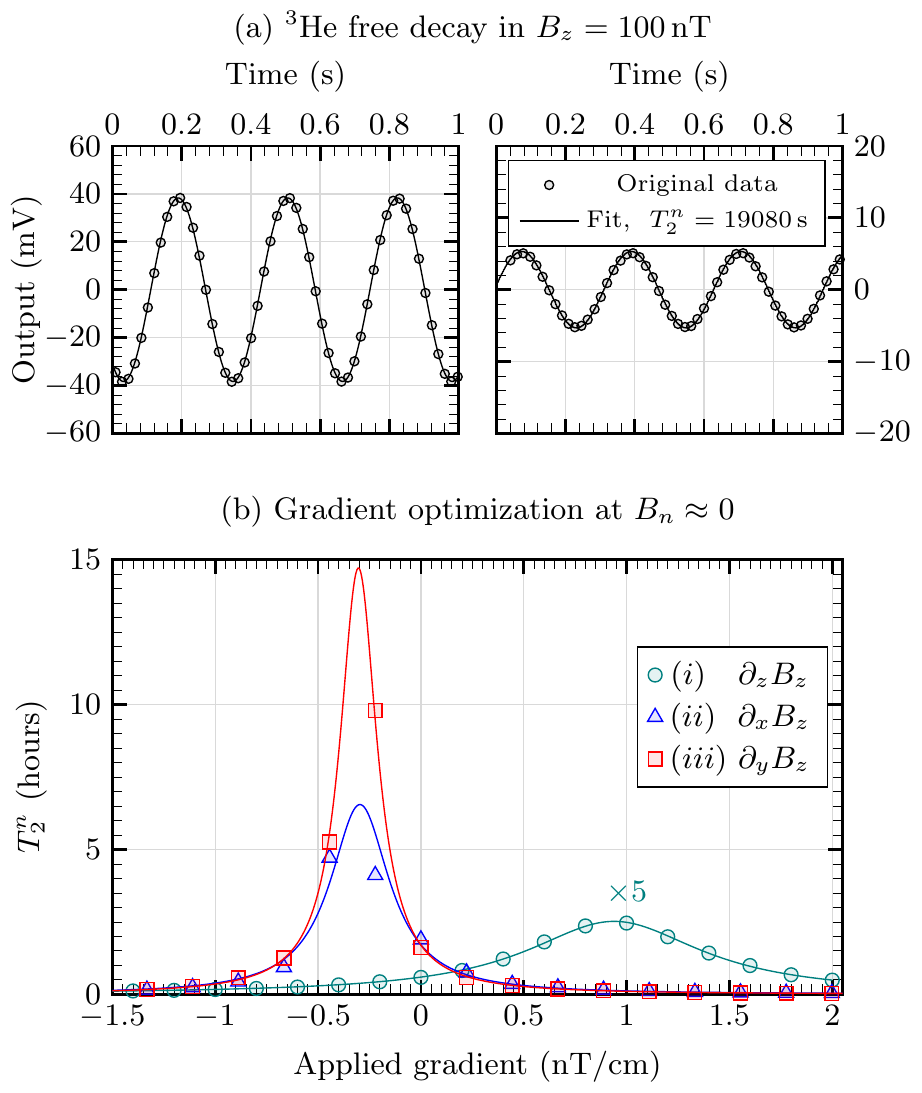}
    \caption{Optimization of the transverse relaxation time of He nuclear spins at low polarization in a leading field of 100\,nT. (a) Samples of $^3$He free decay (left and right panels) separated by a dark time (pump laser off) of 37660\,s (about $10.5$ hours).  Simultaneous fitting of the two data sets leads to $T^n_2 = 19080(40)$\,s  (5.3 hours). (b) Evolution of transverse relaxation time as a function of the $B_z$ magnetic field gradients. The order of optimization goes as $(i)$ $\partial_zB_z$, $(ii)$ $\partial_xB_z$, $(iii)$ $\partial_yB_z$. After a first round of gradient optimization, $T^n_2$ increases from about 200\,s to 15\,hours as suggested by the fits (solid lines), see Eq.\,\eqref{eq:T2LongitudinalGrad}.}
    \label{fig:ini-grad-opti}
\end{figure}

 The procedure is then repeated iteratively for different values of $\partial_iB_z$ gradients, in the sequence $i=z,x,y$. After each sequential step the respective gradient is set to its optimum value. The results are depicted in Fig.\,\ref{fig:ini-grad-opti}; the nuclear spin transverse relaxation time is seen to increase from about 200\,s (no applied field gradients) to 15\,h after a first round of optimization. 
 
Let us note that this method is time-consuming: not only because of the need to destroy $^3$He polarization between measurements, but also because spins have to precess in the dark for some time for a precise measurement of $T^n_2$. 
This becomes increasingly problematic in the course of the optimization procedure as the transverse relaxation time increases. Each point in Fig.\,\ref{fig:ini-grad-opti}(b) necessitates about 10\,min acquisition time.

\begin{figure}[htb]
    \centering
    \includegraphics[scale=0.92]{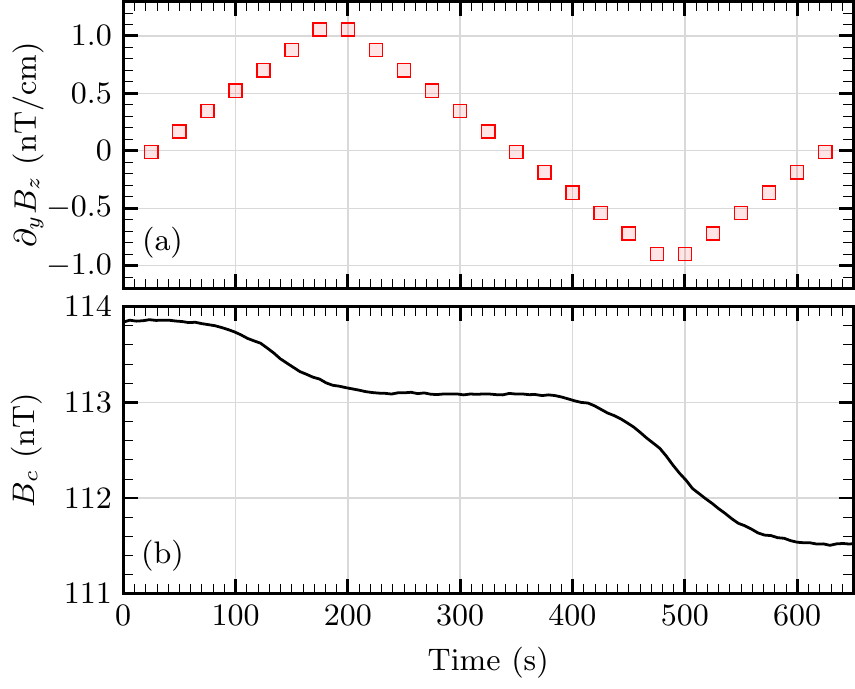}
    \caption{Gradient-zeroing method at the compensation point.  (a) Change of $\partial_yB_z$ gradient value as a function of time. (b) Effect of applied $\partial_yB_z$ gradient on the compensation field as a function of time.}
    \label{fig:comp-point-grad-opti-method}
\end{figure}

\subsubsection{Optimization at the compensation point}\label{sec:opti-comp-point}

Once at the compensation point, we modulate the $B_x$ field with a 40-Hz sine wave of about 0.1\,nT amplitude \footnote{The modulation frequency should be higher than the Larmor frequency of $^3$He, while not being too far away to obtain a reasonable sensitivity.}. 
The response of the comagnetometer with respect to $B_z$ exhibits a dispersive resonance centered at $-B_n$ \cite{lu2020nuclear}, which in our case is $\approx B_c$, see Eq.\,\eqref{eq:simpleCompPoint}. This resonance can be used for closed-loop control of the compensation point based on direct reference to the nuclear polarization which is an alternative to previous approaches locking to electronic resonance  \cite{jiang2019closed,tang2022design}.
Our closed-loop control of the compensation field
allows the comagnetometer to always be operated at optimum sensitivity, and the mechanisms affecting $T^n_1$ or $T^n_2$ can be studied in real time. 

\begin{figure}[htb]
    \centering
    \includegraphics[width=0.48\textwidth]{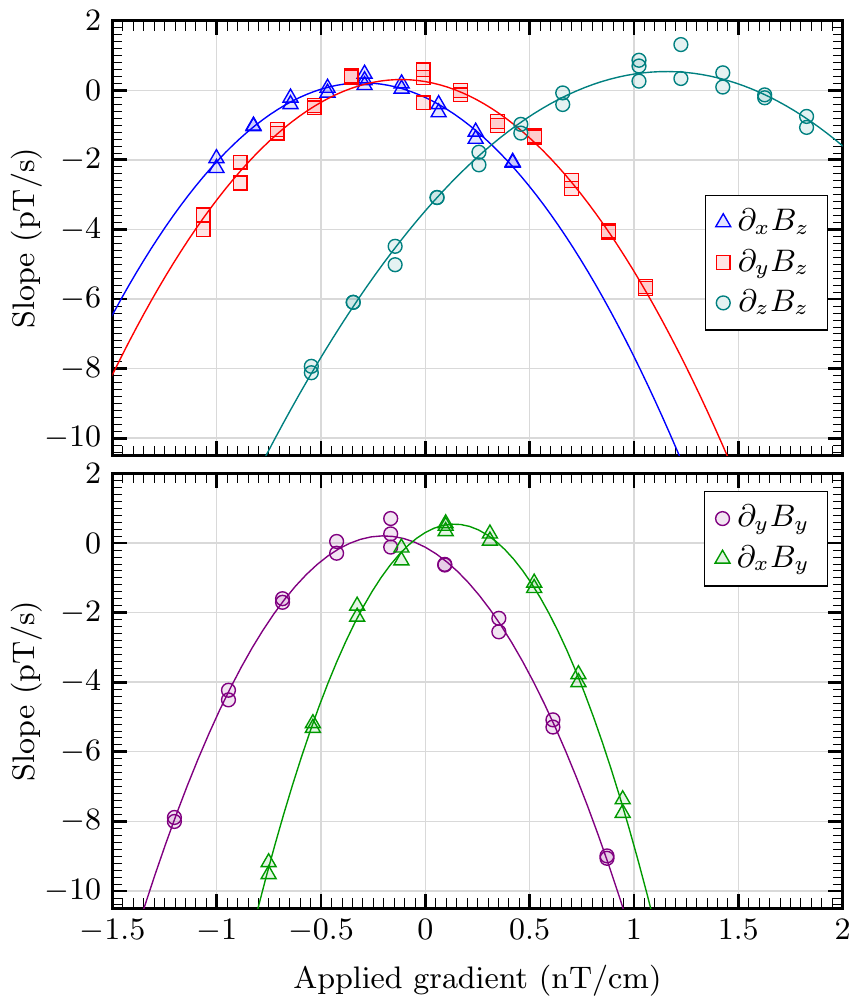}
    \caption{Gradient optimization at the compensation point. The top panel shows the derivative of the time evolution of the compensation point as a function of the applied $B_z$ gradients. The results for the other two independent gradient channels are shown on the bottom panel. The data are fit with a parabola, see Eq.\,\eqref{eq:T1_inh}, for which the center corresponds to the best compensated gradient. For discussions on the slope of parabolas, see the text and Tab.\,\ref{tab:gradientWidth}.}
    \label{fig:comp-point-grad-opti}
\end{figure}

As coils intrinsically generate field gradients, it is important to be able to compensate for them at the working point.  There are five independent first-order gradients, however, the optimization discussed in Sec.\,\ref{sec:T2method}
only involves gradients of the longitudinal field. To maximise $T^n_1$, we use the following routine: we start by locking the field to the compensation point. We then vary one of the currents through the gradient coils over time: values are changed every 25\,s following a saw-tooth pattern, see Fig.\,\ref{fig:comp-point-grad-opti-method}(a). 
This way, the optimum gradient is sampled from different directions at different times, ensuring that effects of drifts are minimized.
As a result of changing the gradient, the equilibrium nuclear polarization is changed, and so is the compensation point, see Fig.\,\ref{fig:comp-point-grad-opti-method}(b), which is observed in real time with the closed-loop control. 

Then, each 25\,s data set corresponding to a field gradient value is differentiated with respect to time in order to obtain the rate of change (in pT/s) of the compensation field as a function of the gradient value. The results are depicted for each of the five independent gradient channels in Fig.\,\ref{fig:comp-point-grad-opti}. Fitting the slope as a function of applied gradient with a parabola (the derivation is given in Appendix\,\ref{sec:AppendII}), the center, which corresponds to the best compensated gradient, is extracted. If the new center differs from the previous known value, we observe a buildup of the compensation field after moving to the new value, which indicates that a higher equilibrium field can be reached. This method has to be applied after large  changes of the stationary compensation field (typically above 10\,nT), since with different leading fields, the gradients change as well. Note that this gradient optimization procedure works also when not at equilibrium nuclear polarization. Indeed, polarization buildup or decay linear in time does not affect the centers of the parabolas shown in Fig.~\ref{fig:comp-point-grad-opti}. If the polarization changes in a nonlinear fashion, this is no longer the case. Thus, it is important to start with close-to-optimum values for the gradients, which can be obtained using the $T^n_2$ method as discussed in Sec.\,\ref{sec:T2method}.

\begin{table}
    \centering
    \def\arraystretch{1.3}
    \begin{tabular}{*4c}
    \toprule
    Coil & Applied $\nabla B_\perp$ gradient & Calc. ratio & Meas. ratio\\
    
    \hline
     \vspace{-0.1cm}
     $\textcolor{green!60!black}{\partial_xB_y}$ & $\partial_xB_y+\partial_yB_x+\,0.34\,\partial_yB_y$  & 2.17 & 2.68(11) \\
      $\textcolor{purple}{\partial_yB_y}$ &$\partial_yB_y-0.68\,\partial_xB_x$ & 1.46 & 1.75(7) \\
$\textcolor{blue}{\partial_xB_z}$ &  $\partial_zB_x$& 1& 1\\
$\textcolor{red}{\partial_yB_z}$ & $\partial_zB_y$ & 1 & 0.95(5)\\
         $\textcolor{teal}{\partial_zB_z}$ &  $-0.5\,\partial_xB_x-0.5\,\partial_yB_y$ & 0.5 & 0.64(3)\\ 
         \bottomrule
    \end{tabular}
    \caption{Calculated and measured slope ratios of the parabolas for each independent magnetic gradient channel. The contribution of applied gradients to the slope is calculated using Eq.\,\eqref{eq:orthoGradComponents}, and compared to that of the $\partial_xB_z$ channel. Each ``Coil'' label refers to the principal component of the generated gradient. Note that, for a given channel, other gradients are also generated with similar or smaller coefficients. The complete set of applied transverse gradients for each channel is listed in the second column. The coefficients depend on the coil windings and are provided by the shield manufacturer. 
    }
    \label{tab:gradientWidth}
\end{table}

Note that the magnetic field gradient generated by a gradient coil affects at least two gradient components. The actual gradient generated in our setup and the expected slope of parabolas in Fig.\,\ref{fig:comp-point-grad-opti} are summarized in Table\,\ref{tab:gradientWidth}. Our results show that, as predicted, the $\partial_xB_y$ gradient is of greatest concern, while the $\partial_zB_z$ has about four times less impact. However, a sizable deviation between expected and measured ratios is also observed. We attribute this error to not well-known calibration factors and coefficients of the gradient coils. Indeed, estimates are provided by the gradient coils manufacturer, obtained by simulating the generated field in the absence of the magnetic shield. This suggest our method for gradient optimization could also be used for \textit{in-situ} calibration of gradient coils which we plan to further investigate.

 \begin{figure}[htb]
    \centering
    \includegraphics[width=0.48\textwidth]{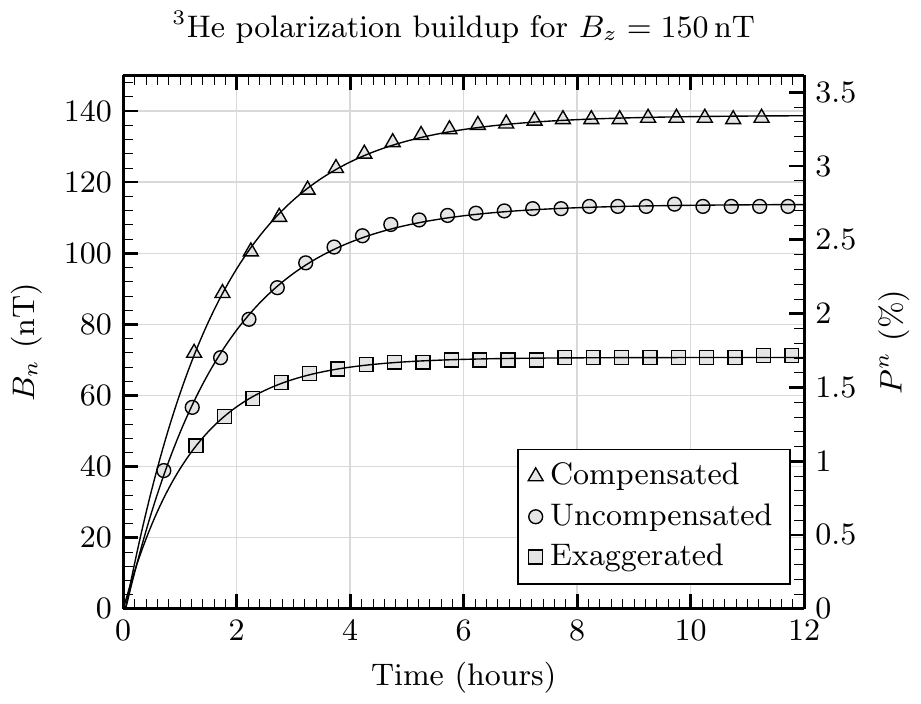}
    \caption{Dynamics of spin-exchange optical pumping of $^{3}$He nuclear spins in a leading field of 150\,nT. The circles are obtained when there was no applied gradients. The squares are obtained when the gradients were exaggerated by changing the sign of the optimized applied gradients (triangles). Each data set was fit (solid lines) assuming an exponential approach of the nuclear field to the equilibrium  value over time, see Eq.\,\eqref{eq:dynamic_pumping}.} 
    \label{fig:HeBuild-up}
\end{figure}

\section{Polarization dynamics}\label{sec:polDyna}

In Fig.\,\ref{fig:HeBuild-up}, we use the experimental method presented in Ref.\,\cite{lu2020nuclear} to show how gradients affect the dynamics of spin-exchange optical pumping of $^3$He nuclear spins for a leading field of 150\,nT and an electronic polarization about 50\%. The electronic polarization is chosen to optimize the sensitivity of the comagnetometer, while the nuclear polarization is determined by the requirement of operation at a stable  compensation point, see Eq.\,\eqref{eq:gradientCondition}. When the gradients are uncompensated the steady-state nuclear field reaches 115\,nT. When the gradients are exaggerated (reversal of all optimized gradients) the steady state value drops to 70\,nT. No stationary compensation point could be found for either of these configurations. 
When the gradients are fully optimized, the nuclear field reaches above $B_{n}\approx 140$\,nT (corresponding to a nuclear polarization of $\approx 3\%$)\,\footnote{Higher nuclear spin polarization can be achieved, for example, $30\%$ in Ref.\,\cite{shaham2022strong}. However, this was done at 100\% electronic polarization and at a higher leading field. Therefore, the $30\%$ polarization is not at the stable compensation point. The polarization in our experiment is similar to that in Ref.\,\cite{vasilakis2009limits}}. In this regime a stable compensation point exists. It was found at 131\,nT, determined by the conditions described by Eq.\,\eqref{eq:gradientCondition}. Note this is slightly below the maximum polarization of 140\,nT achieved with a leading field of 150\,nT displayed in Fig.\,\ref{fig:HeBuild-up}.
At this compensation field, a noise level corresponding to a sensitivity to nuclear spin-dependent energy shifts of $\approx 3 \times 10^{-22}\,\rm{eV}/\sqrt{\rm{Hz}}$ was achieved in the range 0.1 to 1\,Hz. In terms of gyroscopic sensitivity, this corresponds to about $0.5\, \mu\text{rad~s}^{-1}/\sqrt{\text{Hz}}$ ($\approx 80\,\rm{nHz}/\sqrt{\rm{Hz}}$), which is $2.5\,\text{fT}/\sqrt{\text{Hz}}$ in terms of pseudo-magnetic field sensitivity. To estimate the sensitivity of the system we measured the spectrum of the output signal and calibrated it with a  routine based on magnetic field modulation adapted from Ref.\,\cite{kornackPhdThesis2005}.

\begin{figure}[htb]
    \centering
    \includegraphics[width=0.48\textwidth]{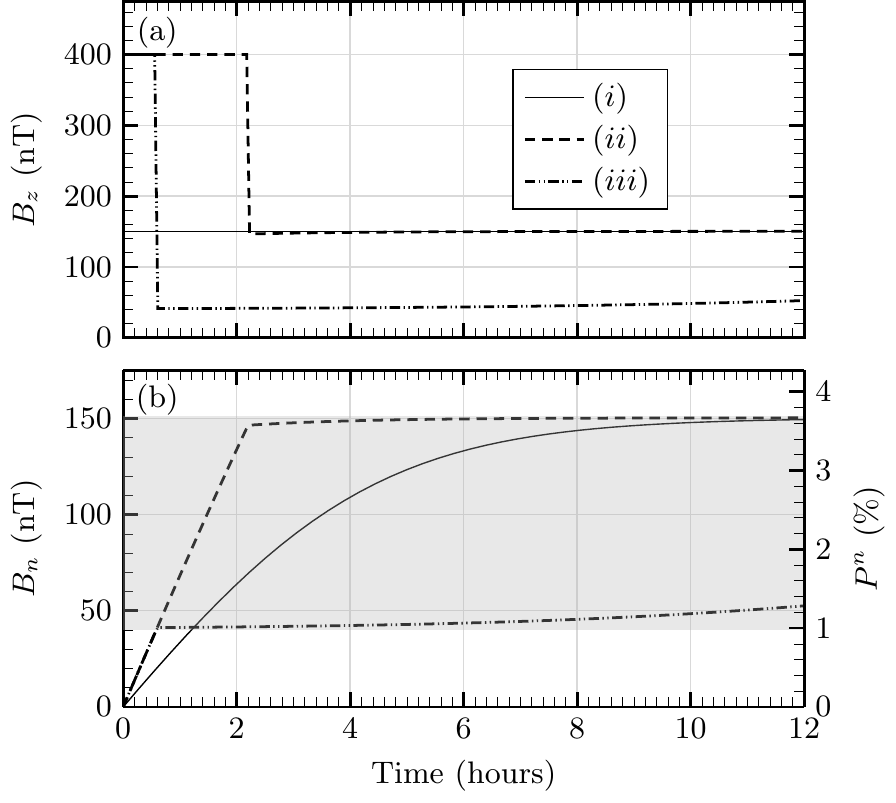}
    \caption{Calculated dynamics of three nuclear polarization strategies. (a) Applied magnetic field as a function of time. (b) Dynamics of the nuclear polarization. Calculations are made with parameters similar to that in Fig.\,\ref{fig:theory} and $|\nabla B_\perp|=0.5\,$nT. The gray shaded area shows nuclear field in between the two stationary compensation points, located at $B_n=40\,$nT and $B_n=150\,$nT, see also Fig.\,\ref{fig:theory}. When the nuclear field is in this area, the closed-loop control of the compensation point can be turned-on and leave the nuclear field move to the highest compensation point over time. Strategy $(i)$ consists in applying a $z$-field equal to the known upper stationary compensation field, $B_n=150\,$nT. Strategies $(ii)$ and $(iii)$ consist in applying, at $t=0$, a larger field than in $(i)$ (here, 400\,nT) and a higher pumping rate (such that $P_e=1$) during the initial stage. Closed-loop control of the compensation field can be turned-on either when reaching the upper stationary compensation field [strategy $(ii)$] or right after passing over the lower one [$B_n=40\,$nT, strategy $(iii)$].} 
    \label{fig:pumpingStrategies}
\end{figure}

Once the system's parameters are determined, different strategies can be employed to reach the stable stationary compensation point starting from an unpolarized system.  The nuclear polarization dynamics for different strategies are illustrated by numerical simulations displayed in Fig.\,\ref{fig:pumpingStrategies}. Considered strategies include:

$(i)$ the leading field is set to the upper stable compensation field value and the optical pumping rate is adjusted to have maximum sensitivity of the comagnetometer, corresponding to $P_e \approx 0.5$;

$(ii)$ the leading field is set to a value much larger than the upper stable compensation field (in our conditions 400\,nT, practically limited by, for example, the coil current source) together with a higher optical pumping rate, such that $P_e \approx 1$, until the upper stationary compensation field is reached. Then the leading field can be locked and the optical pumping rate adjusted such that the system has the highest sensitivity [the conditions of strategy $(i)$];

$(iii)$ the parameters are set to strategy $(ii)$ until the polarization passes the lower stationary compensation point, after which the parameters are set to strategy $(i)$. In this case, the system's nuclear polarization field will continue to grow until it reaches the upper stationary compensation field.

The benefit of the latter strategy is that the comagnetometer is operational in the fastest time, for the given simulation parameters, within 35\,mins. This could be beneficial, for example, for  gyroscopic applications on board vehicles. The changing nuclear polarization while employing strategy $(iii)$ affects the signal-to-noise ratio of the gyroscope, but the device remains operational in the process. Our simulations show, however, that with this strategy, it takes about 50\,hours for the system to reach steady state nuclear polarization. Strategy $(ii)$ is a good alternative to have the system quickly operational at the upper stationary compensation point, i.e. at steady state nuclear polarization. 

\section{Conclusion}
\label{sec:Conclusion}

Motivated by applications in NMR-based rotation sensing, searches for ultralight bosonic dark matter and experiments measuring exotic spin-dependent interactions, we constructed a dual-species (K-$^3$He) comagnetometer with hybrid optical pumping (via Rb atoms) operating in self-compensating regime. 

Building on the body of earlier work by several groups, we developed a method for closed-loop control of the compensation point allowing practical optimization of the operating parameters in spite of the inherently slow (hours to days) dynamics of the system. When the comagnetometer operates at the compensation point, it is generally insensitive to magnetic fields. However, we showed that the system can still be optimized in terms of magnetic gradients and fields without changing the operation mode. 


The presented gradient optimization method facilitates achieving much higher longitudinal and transverse relaxation times as is otherwise possible, improving the polarization level and stability. 
At the compensation point, this also increases the coupling between the nuclear and electronic spins and improves stability of the compensation point.  With these, our method has potential to boost the performance of comagnetometer nuclear-spin-ensemble-based quantum memory \cite{shaham2022strong,katz2022quantum,katz2022optical}, amplifiers \cite{min2021search,min2022Floquet,wang2022limits,wang2022sapphire} and gyroscopes \cite{jiang2019closed,Liu2022Comag,jiang2022single-beam}.

Finally, the device shows a sensitivity comparable to the best previous comagnetometers \cite{vasilakis2009limits}, however, at a several times smaller density of $^3$He. The helium number density of 3\,amg used in this work is lower than in most previous studies. Since lower-pressure cells are easier to manufacture and safer to operate, the optimization technique allows for wide usability of vapor-cell-based nuclear spin sensors in various application areas.

\appendix
\section{Measurement details}
\subsection{SERF magnetometer} \label{sec:appendI-serf}
We measured the polarization of K via Faraday rotation of a linearly polarized probe beam. The  rotation magnitude is in proportional to the component of K polarization $P^e_y$ according to \cite{shah2009spin} 
\begin{equation} \label{eq:faraday_angle}
\theta=\frac{1}{2}n_e r_e c  D f_{D_1} \frac{\nu-\nu_{D_1}}{(\nu-\nu_{D_1})^2+\left[\Gamma^{D_1}_\text{tot}/(2\pi)\right]^2} P^e_y\,,
\end{equation}
where $n_e = 7.5 \times 10^{13}$~cm$^{-3}$ is the number density of K, $r_e = 2.8\times 10^{-15}$\,m is the classical electron radius, $c=3\times 10^{8}$\,m/s is the speed of light, $D=2$\,cm is the cell diameter,  $f_{\text{D}_1}=1/3$ is the oscillator strength for the D$_1$ transition at frequency $\nu_{\text{D}_1}=389.236$\,THz, $\nu=389.539$\,THz is the laser frequency, and $\Gamma^{\text{D}_1}_\text{tot}/(2\pi)=48$\,GHz is the half-width at half-maximum (HWHM) of the transition. With these values, the maximum rotation angle, corresponding to $P^e_y=1$, is 0.7\,rad. 

An optical polarimeter was used to read out the rotation angle as shown in  Fig.\,\ref{fig:setup}, and the decrossing angle between two polarizers was set to $\pi/4$. Thus, the signal $I \propto \sin^2(\theta-\pi/4) \approx 0.5-\theta$ is linearly dependent on the rotation angle.   

To suppress low-frequency drifts, we applied a modulation field $ B_z^m \sin( \omega_m t)$ along the pumping axis (see, for example, Ref.\,\cite{Put2019Nonlinear}). Consequently, $P^e_y$ is a sum of harmonics. Under the assumption that the field component $B_z$ is small,  the approximate solution to the first harmonic is \cite{zhimin2006parametric,jiang2018parametrically}
\begin{equation} \label{eq:soultion_modulation}
\begin{split}
 P^e_y(\omega_m)=&-\frac{2 P^e_z R_\text{tot} \gamma_e Q}{ Q^2R_\text{tot}^2+ (\gamma_e B_y)^2} J_0\left(\frac{\gamma_e B_z^m}{Q \omega_m}\right)\\
 &\times J_1\left(\frac{\gamma_e B_z^m}{Q \omega_m}\right) B_y \sin(\omega_mt)\,,
 \end{split}
\end{equation}
where $J_i(\cdot)$ are Bessel functions of the first kind of order $i$, $Q$ is the slowing-down factor, and  $R_\text{tot}$ is the total relaxation rate of K. For a modulation frequency of 800\,Hz and $Q=5.2$ (assuming $P^e= 50\%$), the factor $ J_0(\frac{\gamma_e B_m}{Q \omega_m}) J_1(\frac{\gamma_e B_z^m}{Q \omega_m})$ is maximized for $B_z^m=161$~nT.  However when $B_z^m$ is large, the modulation field causes rf broadening of magnetic resonance transitions, and the relaxation rate is given by \cite{vasilakis2009limits} 
\begin{equation} \label{eq:soultion_modulation}
R_\text{rf}=\frac{5}{36}\frac{(\gamma_e B_z^m)^2}{Q^2 R_\text{se}}\,,
\end{equation}
where $R_\text{se}=n_e \sigma_{ee} v = 79000$\,s$^{-1}$ is the alkali-alkali spin-exchange rate with the collision cross section $\sigma_{ee} =1.5\times 10^{-14}$\,cm$^2$. For $B_z^m=161$\,nT, the rate is $R_\text{rf}=52$~s$^{-1}$, being 5 times larger than the spin destruction ratio of K on He. We experimentally found that the sensitivity of the K magnetometer peaks at $B_z^m=35$\,nT, which balances the beneficial effect of larger $B_z^m$ on the signal amplitude with its deleterious effect on the linewidth due to the RF broadening. The signal was fed into a lock-in amplifier and demodulated at the first-harmonic to retrieve the rotation angle.

To suppress the alkali polarization gradient, we employ a hybrid pumping technique where an optically thin sample of Rb atoms is optically pumped and the K atoms are polarized via spin-exchange collisions with the Rb atoms. Uniform optical pumping can be achieved with a high ratio of the receiver to donor densities. 

In general, both K and Rb can be the spin donor. Because of the high He pressure in our cell, the spin destruction collisions between alkali atoms and He atoms are the dominant source of relaxation. Given that the spin-destruction cross section of K on He is around 18 times smaller than that of Rb on He \cite{baranga1998,walker2010method}, we chose K as the receiver in the hope of reaching a better magnetometer sensitivity.



Note that the
modulation of $B_x$ or $B_y$ fields also results in a  usable magnetic resonance. However, in that case the projection of the electronic spins on the $z$-axis is then modulated periodically too, leading to a lower average electronic spin polarization along the $z$-axis and therefore a lower equilibrium nuclear polarization, 
in turn affecting the stability of the comagnetometer [Eq.\,\eqref{eq:gradientCondition}].

\subsection{Depolarization of nuclear spins with applied gradients}\label{sec:appendI-depol}

In Sec.\,\ref{sec:T2method}, we proposed a method to optimize the gradients at low nuclear polarization. To make sure all the measurements are performed at the same nuclear polarization, the step $(iv)$ consists of destroying nuclear polarization by applying a 30\,nT/cm $\partial_zB_z$ gradient for 5\,s. It should be stressed that by implementing this step, one needs to make sure that nuclear polarization is completely destroyed. One parameter to consider in this context is the diffusion rate between slices in which the rotation angle of the spins changes by $\pi$. The thickness of such a slice is
\begin{equation}
    \delta z = \frac{1}{2\gamma_n \nabla B_z \delta t}\,,
\end{equation}
where $\gamma_n=3.24\times10^{-2}\,$Hz/nT is the gyromagnetic ratio of $^3$He nuclear spins, $\nabla B_z$ is the applied $z$-gradient and $\delta t$ is the time it is applied.  With $\nabla B_z=30\,$nT/cm and $\delta t = 5\,$s, we find $\delta z=0.11\,$cm. Noting that, for our experimental conditions, $D_{^3\text{He}}\approx 0.71\,$cm$^2$/s, $^3$He spins diffuse through multiple slices during step $(iv)$, which leads to depolarization. Note that, during step $(ii)$, the nuclear magnetization was rotated in the $y-z$ plane. Besides, applying a $\partial_zB_z$ gradient automatically generates a $\partial_yB_y$ gradient, contributing together to the depolarization of longitudinal and transverse nuclear spins [Eqs.\,\eqref{eq:T1_inh} and \eqref{eq:T2LongitudinalGrad}].
\vspace{0.3cm}

\section{Time derivative of polarization in gradient optimization process} \label{sec:AppendII}
The solution of Eq.~\eqref{eq:dynamic_pumping} gives the time evolution of the nuclear polarization of the following form
\begin{equation} \label{eq:nuc_pol_evolv}
\begin{split}
P^n(t)=P^n_s+&[P^n(0)-P^n_s]\\
&\times\left\{1-\exp\left[-(R_{se}^{ne}+1/T^n_1)t\right]\right\}\,, 
\end{split}
\end{equation}
where $P^n(0)$ is an initial value of the polarization and $P^n_s$ is the steady state polarization defined in Eq.~\eqref{eq:nuc_pol} by $P^n(t\rightarrow\infty)$. Therefore, the time derivative of the nuclear polarization is
\begin{equation} \label{eq:nuc_pol_der}
\begin{split}
\frac{d}{dt} P^n(t) = \left[R^{ne}_{\text{se}}+1/T^n_1\right] [P^n_s - P^n(0)] \\ \times\exp[-(R_{\text{se}}^{ne}+1/T^n_1)t]\,.
\end{split}
\end{equation}
In our experiment, the measurement period for a given gradient value lasts for 25\,s, which is much shorter than $T_1^n$, therefore we can approximate $\exp(-t/T_1^n)\approx1$. In addition, $\exp(-R_{se}^{ne}t) \approx 1$ since $R_{\text{se}}^{ne} \lesssim 1/T^n_1$.
Moreover, throughout the whole zeroing procedure the initial polarization for each measurement varies by less than 5\%, therefore for simplicity we assume it to be constant over all measurements within the zeroing procedure.  This leads to the following equation for time derivative of the polarization in each measurement
\begin{equation} \label{eq:nuc_pol_der}
\begin{split}
\frac{d }{dt}P^n(t) &= (R_{\text{se}}^{ne}+1/T^n_1)(P^n_s-P^n(0))\,,
\end{split}
\end{equation}
then using Eqs.~\eqref{eq:nuc_pol} and \eqref{eq:T1_inh}, one can get an expression of polarization time derivative as a function of the gradients of transversal fields $|\nabla B_\perp|$ (see Fig.\,\ref{fig:comp-point-grad-opti})
\begin{equation}
      \frac{d }{dt}P^n(t) = R_{\text{se}}^{ne}[P^e-P^n(0)]-\frac{D_{n}}{B_z^2}P^n(0) |\nabla B_\perp(t)|^2\,, 
\end{equation}
where time dependence of the gradient is due to the change of its value during the zeroing routine.

\section*{Acknowledgements}
We thank Wei Ji for fruitful discussions. This work was supported by the German Federal Ministry of Education and Research (BMBF) within the Quantentechnologien program (FKZ 13N15064). The work of D.F.J.K. was supported by U.S. National Science Foundation (NSF) Grant No. PHY-2110388. SP and MP acknowledge support by the National Science Centre, Poland within the OPUS program (2020/39/B/ST2/01524).
\bibliography{Gyroliterature}

\clearpage

\end{document}